# BRIGHT FIREBALLS ASSOCIATED WITH THE POTENTIALLY HAZARDOUS ASTEROID 2007LQ19

**José M. Madiedo[1], Josep M. Trigo-Rodríguez[2], José L. Ortiz[3], Alberto J. Castro-Tirado[3] and Jesús Cabrera-Caño[1]**

[1] Departamento de Física Atómica, Molecular y Nuclear. Facultad de Física. Universidad de Sevilla. 41012 Sevilla, Spain.

[2] Institute of Space Sciences (CSIC-IEEC), Campus UAB, Facultat de Ciències, Torre C5-parell-2ª, 08193 Bellaterra, Barcelona, Spain.

[3] Instituto de Astrofísica de Andalucía, CSIC, Apt. 3004, 18080 Granada, Spain.

**ABSTRACT**

We analyze here two very bright fireballs produced by the ablation in the atmosphere of two large meteoroids in 2009 and 2010. These slow-moving and deep-penetrating events were observed over Spain in the framework of our Spanish Fireball Network continuous meteor monitoring campaign. The analysis of the emission spectrum imaged for one of these fireballs has provided the first clues about the chemical nature of the progenitor meteoroids. The orbital parameters of these particles suggests a likely association with the recently identified July ρ-Herculid meteoroid stream. In addition, considerations about the likely parent body of this stream are also made on the basis of orbital dissimilarity criteria. This orbital analysis reveals that both meteoroids and PHA 2007LQ19 exhibit a similar evolution during a time period of almost 8,000 years, which suggests that either this NEO is the potential parent of these particles or that this NEO and both meteoroids had a common progenitor in the past.

## 1 INTRODUCTION

The development of asteroid surveys has resulted in a vast increase in the number of known near Earth objects (NEOs), some of which could be dynamically evolved Jupiter Family Comets (JFCs) (Jewitt 2008, Morbidelli 2008, Williams 2009). Many of these inactive (i.e. dormant) comets are formed by very dark materials, with geometric albedos well below 0.1 (Trigo-Rodríguez et al. 2014). Some of these inactive comets are a potential source of impact hazard and could be wrongly classified as Near Earth Asteroids (NEAs) in absence of dynamic clues (Williams 2009). In this context it is crucial for future hazard mitigation programs to know the real nature of these bodies. NEO monitoring programs are trying to catalogue all these objects and have led, in turn, to a corresponding increase in the number of claimed associations between meteoroid streams and these objects. A review about the relationships between meteoroids and their parents can be found in Williams (2011). The standard way of linking meteoroids with a given comet or asteroid is through the similarity of their orbits. This technique is based on the calculation of the so-called dissimilarity function, which measures the "distance" between the orbit of the meteoroid and that of the potential parent. If this distance is below a predefined cut-off value, orbits are deemed to be similar. Porubčan et al. (2004) studied the similarity of meteoroid orbits with their parent bodies for a period of at least 5,000 years before a generic association is claimed, a view later supported by Trigo-Rodriguez et al. (2007). The first dissimilarity criterion was proposed by Southworth and Hawkins (1963), who defined the $D_{SH}$ dissimilarity function for two orbits A and B:

$$D_{SH}^2 = (e_B - e_A)^2 + (q_B - q_A)^2 + \left(2\sin\frac{I_{BA}}{2}\right)^2 + \left(\frac{e_A + e_B}{2}\right)^2 \left(2\sin\frac{\pi_{BA}}{2}\right)^2 \qquad (1)$$

In this equation q and e are, respectively, the perihelion distance (expressed in astronomical units) and the orbital eccentricity. $I_{BA}$ is the angle between the two orbital planes and $\pi_{BA}$ the difference between the longitudes of perihelia as measured from the intersection of both orbits. A small value of $D_{SH}$ implies a small difference between both orbits, which can then be regarded as similar. Usually, a cut-off value of 0.15 is adopted for $D_{SH}$ (Lindblad 1971a and 1971b). Alternative dissimilarity criteria were defined later on by Drummond (1981), Jopek (1993), Valsecchi et al. (1999) and Jenniskens (2008).

Recently, by using data obtained from the systematic monitoring of fireballs in the atmosphere, this approach has been employed by Madiedo et al. (2013) to establish a link between the potentially hazardous asteroid 2008XM1 and the Northern χ-Orionid meteoroid stream. Using a similar methodology it has also been established that NEO 2012XJ112 is a likely source of large achondritic meteoroids impacting our planet (Madiedo et al. 2014). Additional NEA associations with meteoroid complexes have been established by other authors (Babadzhanov et al. 2009, 2012, 2013). Thus, a systematic and long-term monitoring of meteor and fireball activity demonstrates its potential to establish dynamic associations between meteoroids and NEOs. In this context, we analyze here two bright and slow-moving fireballs recorded over the South of Spain in July 2009 and 2010. Their atmospheric trajectory, radiant, and orbital parameters are calculated, and their likely parent body is identified. Besides, the emission spectrum imaged for one of these events is discussed.

## 2 INSTRUMENTATION AND DATA REDUCTION TECHNIQUES

To image the fireballs discussed here we employed an array of CCD video devices operating from the meteor observing stations listed in Table 1. These employed between 5 and 12 high-sensitivity monochrome Watec cameras (models 902H and 902H Ultimate, from Watec Corporation, Japan) to monitor the night sky. The video cameras are equipped with a 1/2" Sony interline transfer CCD image sensor with their minimum lux rating ranging from 0.01 to 0.0001 lux at f1.4. These cameras generate interlaced imagery according to the PAL video standard, at a rate of 25 frames per second and with a resolution of 720x576 pixels. The observing stations work in an autonomous way by means of the MetControl software (Madiedo et al. 2010). One of the tasks that this software performs is the identification of meteor trails simultaneously imaged from two different observing stations. The images of each multi-station event are then analyzed in order to obtain the atmospheric trajectory of the meteor and the orbital parameters of the progenitor meteoroid. For this data reduction we first perform an astrometric measurement by hand in order to obtain the plate (x, y) coordinates of the meteor along its apparent path from each station. The astrometric measurements are then introduced in our Amalthea software (Trigo-Rodríguez et al. 2009; Madiedo et al. 2011a), which transforms plate coordinates into equatorial coordinates by using the position of reference stars appearing in the images. This package employs the method of the intersection of planes to determine the atmospheric trajectory of the meteor and the position of the apparent radiant (Ceplecha 1987). The preatmospheric velocity $V_\infty$ is found by measuring the velocities at the earliest part of the meteor trajectory. Once these data are known, the software computes the orbital elements of the parent meteoroid by following the standard technique described in Ceplecha (1987).

On the other hand, the evolution of the magnitude of the fireballs analyzed here along their trajectory was determined by direct comparison of the brightness level of the pixels of the meteor trail and those of reference stars appearing in the images. However, at those stages where this comparison was not possible because the brightness of the event was higher than that of reference stars, a previous calibration of our cameras with bright sources in a laboratory was employed to obtain the corresponding magnitudes. In this way, the visual magnitudes were determined with an uncertainty of about 0.5 magnitudes.

In addition, in order to obtain information about the chemical nature of meteoroids ablating in the atmosphere, a systematic spectroscopic monitoring was performed at stations #1, 3 and 4. For this purpose we employed holographic diffraction gratings (500 or 1000 lines/mm, depending on the device) attached to the objective lens of some of the above-mentioned video cameras. In this way, these CCD devices were configured as slitless video spectrographs.

## 3 OBSERVATIONS AND RESULTS

The bolides analyzed here were included in our fireball database with SPMN codes 050709 and 040710. As usual, these codes were assigned after the recording date with format 'ddmmyy', with d being the observing day, m the month and y the year. Thus, these events were imaged on 5 July 2009 at 4h15m29.4 ± 0.1s UTC and 4 July 2010 at 23h16m01.9 ± 0.1s UTC, respectively.

The SPMN050709 fireball (Figure 1) was simultaneously recorded from stations #1, 2 and #4 in Table 1. We named this event "Golfo de Cádiz", after the name of the area in the ocean it overflew. The analysis of the images revealed that the meteoroid impacted the atmosphere with an initial velocity $V_\infty = 17.7 \pm 0.3$ km s$^{-1}$, and the zenith angle of the atmospheric trajectory was of about 69.4º. The apparent radiant was located at the equatorial coordinates $\alpha = 259.7 \pm 0.3º$, $\delta = 30.8 \pm 0.1º$. The luminous phase began at a height of 90.4 ± 0.5 km above the sea level. The fireball exhibited a sudden increase in brightness at a height of 61 ± 1 km, reaching an absolute magnitude of -9.0 ± 0.5 during its brightest phase. Unfortunately, the final part of the luminous trajectory was not imaged. Thus, the bolide disappeared from the field of view of the cameras when it was located at a height of 43.9 ± 0.5 km above the sea level, and the position of the terminal point, which of course was below this level, could not be established.

On the other hand, the SPMN040710 bolide was recorded from stations #1 and 3. We named it "Chipiona", after the name of the coastal town in the South of Spain whose zenith was near from the brightest phase of this fireball. The event is shown in Figure 2 at four different phases along its atmospheric path. This fireball exhibited a very bright flare and reached an absolute magnitude of -14.5 ± 0.5 at its maximum phase. Due to its extraordinary brightness, numerous casual eyewitnesses in the south of Spain reported this event by means of our website (www.spmn.uji.es). However, these visual reports were not accurate enough and so they were not included in the computations. Our calculations show that the meteoroid struck the atmosphere with an initial velocity $V_\infty$ = 18.5 ± 0.4 km s$^{-1}$. The angle of the atmospheric path to the Earth's surface was of about 10.9°. The apparent radiant was located at $\alpha$ = 256.4 ± 0.1°, $\delta$ = 30.8 ± 0.1°. The fireball began at a height of 96.6 ± 0.5 km and ended at 42.1 ± 0.5 km over the sea level.

The main parameters of their luminous path of both events are summarized in Table 2. The beginning ($H_b$), ending ($H_e$) and maximum brightness ($H_m$) heights are indicated, together with the absolute magnitude (M) of the bolides. As shown in Figure 3, in both cases the atmospheric trajectory was located over the Atlantic Ocean. The orbital elements of the parent meteoroids are summarized in Table 3, where a is the semi-major axis, e the orbital eccentricity, i the inclination, $\omega$ the argument of perihelion, $\Omega$ the longitude of the ascending node and q the perihelion distance. The orbital period P and the Tisserand parameter with respect to Jupiter ($T_J$) are also indicated.

## 4 DISCUSSION

### 4.1. Orbital parameters

The values obtained for the radiant position, geocentric velocity and orbital parameters show that both fireballs are dynamically related. The only currently known meteoroid stream that fits these data is the recently identified July $\rho$-Herculids (JRH), whose orbital parameters are shown in Table 4 (Rudawska & Jenniskens, 2014; Kornoš et al. 2014). This poorly-known meteoroid swarm is currently included in the IAU working list of meteor showers. Nevertheless, the JRH stream was identified from the analysis of meteors imaged between July 24 and August 1 (Rudawska, personal comm.), about 20 days after the events analyzed here were observed. But the activity period of this stream is not well-established yet, and this time difference could explain the discrepancies between the averaged position of the geocentric radiant corresponding to these bolides ($\alpha_g = 254.4 \pm 0.4^\circ$, $\delta_g = 27.9 \pm 0.2^\circ$) and the geocentric radiant shown for the JRH in Table 4. Thus, this would imply a radiant drift of about 0.6 degrees per day in right ascension and 0.4 degrees per day in declination. But, however, since no JRH meteors were imaged between both dates, the likely association between these two bolides and the JRH stream cannot be unambiguously established. Additional observations of JRH meteors would be necessary in order to obtain more accurate orbital elements, more information about its complete activity period, and a better understanding of the dynamical evolution of this stream.

On the other hand, the Tisserand parameter with respect to Jupiter obtained from the orbital data in Table 3 yields $T_J = 3.0 \pm 0.1$ and $T_J = 3.0 \pm 0.1$ for the SPMN050709 and SPMN040710 meteoroids, respectively. Thus, from the point of view of this parameter, before impacting our planet the meteoroids were following an orbit which can be considered as a limiting case between that of an asteroid and that of a Jupiter family comet (JFC). Besides, within the observational uncertainty, the calculated orbital periods shown in Table 3 reveal that these meteoroids were trapped in a 3:0 orbital resonance with Jupiter.

### 4.2 Lightcurves and preatmospheric masses

The lightcurves obtained from the photometric analysis of the images obtained for both events are shown in Figures 4 and 5. These reveal that both bolides exhibited their maximum luminosity during the second half of their atmospheric trajectory. The initial (preatmospheric) photometric mass of both meteoroids have been estimated as the total mass lost due to the ablation process between the beginning of the luminous phase and the terminal point of the atmospheric trajectory:

$$m_p = 2 \int_{t_e}^{t_b} I_p /(\tau v^2) dt \qquad\qquad (2)$$

where $m_p$ is the photometric mass, $t_b$ and $t_e$ are, respectively, the times corresponding to the beginning and the end of the luminous phase and $I_p$ is the measured luminosity of the fireball, which is related to the absolute magnitude M by means of the following relationship:

$$M = -2.5 \log_{10} (I_p) \qquad\qquad (3)$$

In equation (2), the luminous efficiency $\tau$ has been estimated from the equation provided by Ceplecha & McCrosky (1976) for velocities v ranging between 17 and 27 km s$^{-1}$:

$$\log (\tau) = -12.50 + 0.17 \log_{10}(v) \qquad\qquad (4)$$

It must be taken into account that, since the final part of the luminous trajectory of the SPMN050709 "Golfo de Cádiz" event was not imaged, the integration in equation (2) was performed for this bolide between the beginning of its luminous phase and the end of the recordings obtained by our cameras. So, the value obtained from this analysis should be considered as a lower limit for the preatmospheric mass of the SPMN050709 meteoroid. Thus, the calculations yield an initial mass of $23 \pm 2$ and $280 \pm 25$ kg for the SPMN050709 and the SPMN040710 meteoroids, respectively. By assuming a spherical shape, the diameter of these particles would be, respectively, $26.3 \pm 0.7$ and $60.6 \pm 2.0$ cm for a bulk density $\rho_m = 2.4$ g cm$^{-3}$. For a bulk density $\rho_m = 3.7$ g cm$^{-3}$ the calculated diameter yields $22.8 \pm 0.6$ cm for the SPMN050709 meteoroid and $52.4 \pm 2.8$ cm for the SPMN040710 meteoroid.

## 4.3 Tensile strength

As can be noticed in the composite images shown in Figures 1 to 2 and also in the lightcurves plotted in Figures 4 and 5, both fireballs exhibited a very bright flare along their atmospheric path. These events are typically produced by the fragmentation of the meteoroids when these particles penetrate denser atmospheric regions. The following equation provides the value of the aerodynamic strength S at which these breakups take place (Bronshten 1981):

$$S = \rho_{atm} \cdot v^2 \tag{5}$$

where v is the velocity of the meteoroid at the disruption point and $\rho_{atm}$ the atmospheric density at the height where the break-up occurs. This aerodynamic strength S can be used as an estimation of the tensile strength of the meteoroid (Trigo-Rodriguez & Llorca 2006, 2007). Here we have employed the standard atmosphere model (U.S. Standard Atmosphere 1976) to obtain the atmospheric density.

The SPMN050709 fireball exhibited a bright flare at a height of 61 ± 1 km above the sea level, when meteor velocity was 17.3 ± 0.5 km s$^{-1}$. According to equation (5), the aerodynamic strengths at which this flare took place was $(7.5 \pm 1.5) \times 10^5$ dyn cm$^{-2}$. On the other hand, the catastrophic disruption experienced by the progenitor meteoroid of the SPMN040710 bolide took place at a height of 65 ± 1 km, when the velocity was 17.9 ± 0.4 km s$^{-1}$. So, this breakup took place under an aerodynamic strength of $(4.8 \pm 0.9) \times 10^5$ dyn cm$^{-2}$. A part of the material, however, was observed to penetrate the atmosphere beyond this point and, as mentioned above, the luminous phase finished at a height of 42.1 km above the sea level. For both meteoroids, the calculated value for the tensile strength is above, but not too different, from the strength of $(3.4 \pm 0.7) \times 10^5$ dyn cm$^{-2}$ exhibited by tough meteoroids with a cometary origin (Trigo-Rodriguez & Llorca 2006, 2007).

### 4.4 Emission spectrum of the SPMN050709 fireball

The emission spectrum of the SPMN050709 bolide was recorded by a videospectrograph operating at station #4 in Table 1. The identification of the main lines appearing in it has been performed with the CHIMET software (Madiedo et al. 2011b). The video file containing the spectrum was first deinterlaced, and then the video frames were sky-background-substracted and flat-fielded. Next, by using typical emission lines appearing in meteor spectra, the signal was calibrated in wavelength and then corrected by means of the instrumental efficiency of the recording device. The result is plotted in Figure 6, which also shows the main lines identified in the signal.

Multiplet numbers are given according to Moore (1945). As usual in meteor spectra, most lines correspond to Fe-I multiplets that are widely distributed along the signal (see e.g. Trigo-Rodríguez et al., 2003). The most significant contributions correspond to Fe I-5 (373.7 nm), Fe I-43 (414.3 nm), the Mg I-2 triplet at 516.7 nm and the Na I-1 doublet at 588.9 nm. The contributions from multiplets Ca I-2 (422.6 nm) and Fe I-41 (441.5 nm) were also identified.

To obtain information about the likely nature of the meteoroid we have analyzed the relative intensity of the emission lines from Na I-1, Mg I-2 and Fe I-15 multiplets (Borovička et al. 2005). Thus, their intensity was measured frame by frame in the corresponding video containing the spectrum and corrected by taking into account the spectral sensitivity of the spectrograph. Then, the contributions in each frame were added to obtain the integrated intensity for the above-mentioned multiplets along the atmospheric trajectory of each fireball. In this way, the Na/Mg and Fe/Mg intensity ratios yield 1.00 and 1.13, respectively. As shown in Figure 5 in Borovička et al. (2005), the estimated value of the Na/Mg intensity ratio fits fairly well the result expected for meteoroids with chondritic composition when the meteor velocity is of about 18 km s$^{-1}$. Besides, we have plotted the Na I-1, Mg I-2 and Fe I-15 relative intensities in a ternary diagram (Figure 7). The solid curve in this graph represents the expected relative intensity, as a function of meteor velocity, for chondritic meteoroids (Borovička et al. 2005). This plot shows that, according to the classification given by Borovička et al. (2005), the emission spectrum recorded for the "Golfo de Cádiz" fireball can be considered as normal, since the point describing this bolide fits fairly well the expected relative intensity for a meteor velocity of about 18 km s$^{-1}$. So, our results suggest a chondritic nature for the meteoroid.

### 4.5 Parent body

In order to establish a link between the fireballs and their potential parent body we have employed the ORAS software, which is described in detail in Madiedo et al. (2013). This application searches through the NeoDys (http://newton.dm.unipi.it/neodys/) and the Minor Planet Center (http://www.minorplanetcenter.org/iau/mpc.html) databases in order to establish a potential similarity between the orbit of meteoroids and the orbits of other bodies in the Solar System using dissimilarity criteria. In this case, we have employed the $D_{SH}$ dissimilarity function proposed by Southworth & Hawkins (1963).

We have chosen the usual cut-off value of 0.15 adopted for $D_{SH}$ (Lindblad 1971a and 1971b). The computations were performed with the averaged orbit of the progenitor meteoroids of the SPMN050709 and SPMN040710 bolides (Table 3). As a result of this analysis, only one potential parent body satisfying the condition $D_{SH} \leq 0.15$ was found. Thus, a value of $D_{SH}=0.07$ was obtained for NEO 2007LQ19. The orbital elements of this potentially hazardous asteroid (PHA) are also shown in Table 3. These orbital data were taken from the JPL database (http://ssd.jpl.nasa.gov/). Then, the Mercury 6 symplectic integrator (Chambers 1999) was employed to perform a numerical integration backwards in time for 10,000 years of the averaged orbit of the fireballs and this NEO in order to determine whether their orbital evolution was similar. In this computation we took into account, in addition to the Sun, the gravitational fields of Venus, Earth, Moon, Mars, Jupiter and Saturn. Figure 8 shows that the values of the $D_{SH}$ dissimilarity function remain below or equal to the cut-off value of 0.15 during almost 8,000 years, a period of time larger than the 5,000 years time scale required by Porubčan et al. (2004). This result strongly suggests a link between the fireballs and this PHA. In fact, as can be seen in Figure 9, the evolution over time of the orbital elements of 2007LQ19 and the meteoroid is very similar. Despite this analysis shows that at present the best fit to the averaged orbit of both meteoroids is 2007LQ19, there could of course be an as yet undiscovered NEO that could provide a better fit. This was, for instance, the case for the Northern χ-Orionids meteoroid stream, where initially 2002XM35 was thought to be the parent (Porubčan et al. (2004) but after discovery 2008XM1 proved to provide a closer fit (Madiedo et al. 2013), suggesting also that both 2008XM1 and 2002XM35 had a common parent body in the past. So, 2007LQ19 and both meteoroids could be a family of objects evolving very similarly and produced by a common parent body.

## 5 CONCLUSIONS

We have analyzed two fireballs observed over Spain and catalogued as SPMN050709 and SPMN040710. The main conclusions derived from this work are as follows:

1) Both fireballs were deep-penetrating and slow-moving events. The photometric data show that the progenitor meteoroids of the mag. -9.0 ± 0.5 SPMN050709 fireball and the mag. -14.5 ± 0.5 SPMN040710 bolide had an initial mass of 23 ± 2 and 280 ± 25 kg, respectively. The SPMN050709 fireball penetrated the atmosphere till a final height below 43.9 ± 0.5 km, and the SPMN040710 event reached the terminal point of its luminous path at a height of 42.1 ± 0.5 km above the sea level.

2) The calculated radiant and orbital parameters show that both meteoroids were dynamically related. Within the observational uncertainty, both particles were found to be trapped in a 3:1 resonance with Jupiter. Besides, these data suggest that both particles could be linked to the recently identified July ρ-Herculid meteoroid stream. However, additional observations of meteors from this poorly-known stream would be necessary in order to confirm this association.

3) The tensile strength calculated for the meteoroids is above, but not too different, from the strength exhibited by tough meteoroids with a cometary origin.

4) The orbital analysis performed with our orbital association software by employing the Southworth and Hawkins dissimilarity function suggests that the potentially hazardous asteroid 2007LQ19 is the parent body of both meteoroids. Thus, the orbit of this NEO and the averaged orbit of both particles remain similar ($D_{SH} \leq 0.15$) over a time period of almost 8,000 years.

5) The analysis of the emission spectrum imaged for the SPMN050709 fireball indicates a chondritic nature for these meteoroids.

## ACKNOWLEDGEMENTS

We acknowledge support from the Spanish Ministry of Science and Innovation (project AYA2011-26522) and Junta de Andalucía (project P09-FQM-4555).

**TABLES**

Table 1. Geographical coordinates of the meteor observing stations involved in this work.

| Station # | Station name | Longitude (W) | Latitude (N) | Alt. (m) |
|:---:|:---:|:---:|:---:|:---:|
| 1 | Sevilla | 5º 58' 50" | 37º 20' 46" | 28 |
| 2 | Huelva | 6º 56' 11" | 37º 15' 10" | 25 |
| 3 | El Arenosillo | 6º 43' 58" | 37º 06' 16" | 40 |
| 4 | Doñana | 6º 34' 06" | 37º 00' 58" | 20 |

Table 2. Trajectory and radiant data for the bolides analyzed in the text. M is the peak absolute magnitude, $H_b$, $H_m$ and $H_e$ the beginning, maximum brightness and ending heights of the luminous trajectory, respectively. $\alpha_g$ and $\delta_g$ are the right ascension and declination of the geocentric radiant (J2000). $V_\infty$, $V_g$ and $V_h$ indicate the observed (preatmospheric), geocentric and heliocentric velocity, respectively.

| SPMN Code | Date | Time (UTC) ±0.1s | M | $H_b$ (km) | $H_m$ (km) | $H_e$ (km) | $\alpha_g$ (º) | $\delta_g$ (º) | $V_\infty$ (km s$^{-1}$) | $V_g$ (km s$^{-1}$) | $V_h$ (km s$^{-1}$) |
|:---:|:---:|:---:|:---:|:---:|:---:|:---:|:---:|:---:|:---:|:---:|:---:|
| 050709 | July 5, 2009 | 04h15m29.4s | -9.0 ±0.5 | 90.4 ±0.5 | 61 ±1 | <43.9 ±0.5 | 250.9 ±0.4 | 25.6 ±0.2 | 17.7 ±0.3 | 14.1 ±0.4 | 37.6 ±0.4 |
| 040710 | July 4, 2010 | 23h16m01.9s | -14.5 ±0.5 | 99.6 ±0.5 | 65 ±1 | 42.1 ±0.5 | 253.8 ±0.1 | 30.1 ±0.1 | 18.5 ±0.4 | 14.8 ±0.5 | 37.3 ±0.4 |

Table 3. Orbital elements (J2000) for both fireballs and for NEO 2007LQ19. The averaged orbit of both fireballs is also shown.

| Object | a (AU) | e | i (º) | Ω (º) | ω (º) | q (AU) | P (yr) | $T_J$ |
|:---:|:---:|:---:|:---:|:---:|:---:|:---:|:---:|:---:|
| SPMN050709 | 2.68±0.12 | 0.640±0.018 | 16.4±0.5 | 103.1582±10$^{-4}$ | 209.4±0.2 | 0.965±0.001 | 4.4±0.3 | 3.0±0.1 |
| SPMN040710 | 2.51±0.17 | 0.617±0.027 | 18.7±0.5 | 102.7158±10$^{-4}$ | 210.0±0.1 | 0.964±0.001 | 4.0±0.4 | 3.1±0.1 |
| Averaged | 2.59±0.15 | 0.628±0.023 | 17.5±0.5 | 102.9370±10$^{-4}$ | 209.7±0.2 | 0.964±0.001 | 4.2±0.4 | 3.0±0.1 |
| 2007LQ19 | 2.60 | 0.629 | 17.0 | 110.8921 | 207.5 | 0.965 | 4.2 | 3.0 |

Table 4. Averaged geocentric velocity $V_g$, orbital elements and geocentric radiant position ($\alpha_g$, $\delta_g$) for the July ρ-Herculids according to Kornoš et al. (2014) and Rudawska & Jenniskens (2014) for N=8 meteors (J2000).

| | $\alpha_g$ (º) | $\delta_g$ (º) | $V_g$ (km s⁻¹) | a (AU) | e | i (º) | $\Omega$ (º) | $\omega$ (º) | q (AU) |
|---|---|---|---|---|---|---|---|---|---|
| Rudawska & Jenniskens | 265.1 | 36.4 | 15.6 | 2.67 | 0.633 | 21.3 | 124.6 | 203.8 | 0.981 |
| Kornoš et al. | 265.9±2.8 | 36.2±2.3 | 14.18±0.97 | 2.20±0.07 | 0.553±0.046 | 19.7±1.5 | 125.8±5.2 | 204.5±5.4 | 0.982±0.012 |

<u>FIGURES</u>

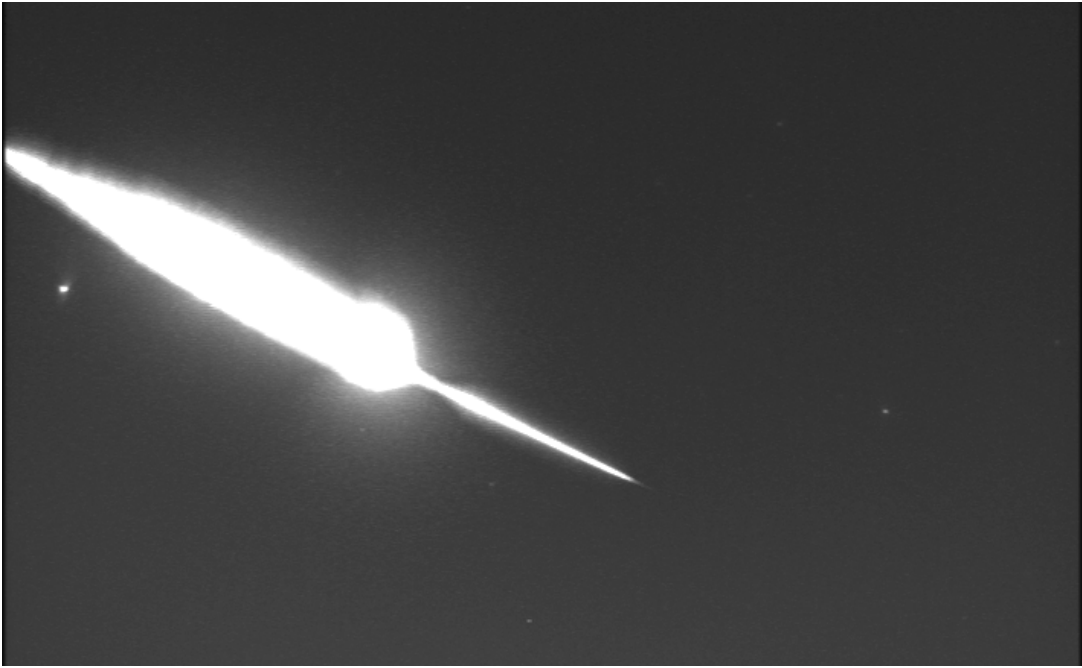

Figure 1. Composite image of the SPMN050709 "Golfo de Cádiz" fireball imaged on
July 5, 2009 at 4h15m29.4 ± 0.1s UTC from station #4 (Doñana).

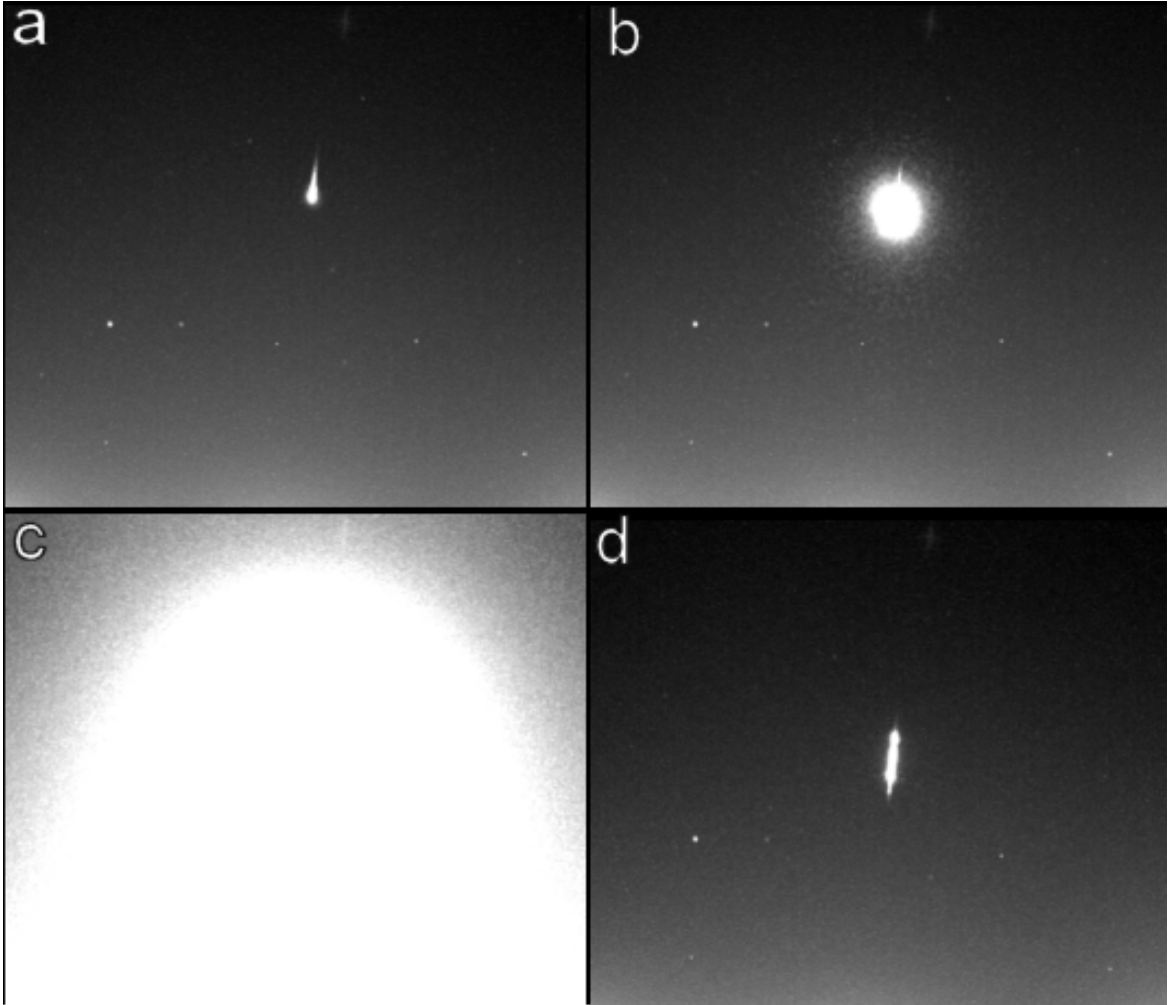

Figure 2. Images of the SPMN040710 "Chipiona" bolide imaged on July 4, 2010 at 23h16m01.9 ± 0.1s UTC from Sevilla: a) and b) show the initial instants of the luminous phase (at t = 1 s and t = 1.5 s after the beginning of the event, respectively); c) main flare (brightest phase); d) persistent train.

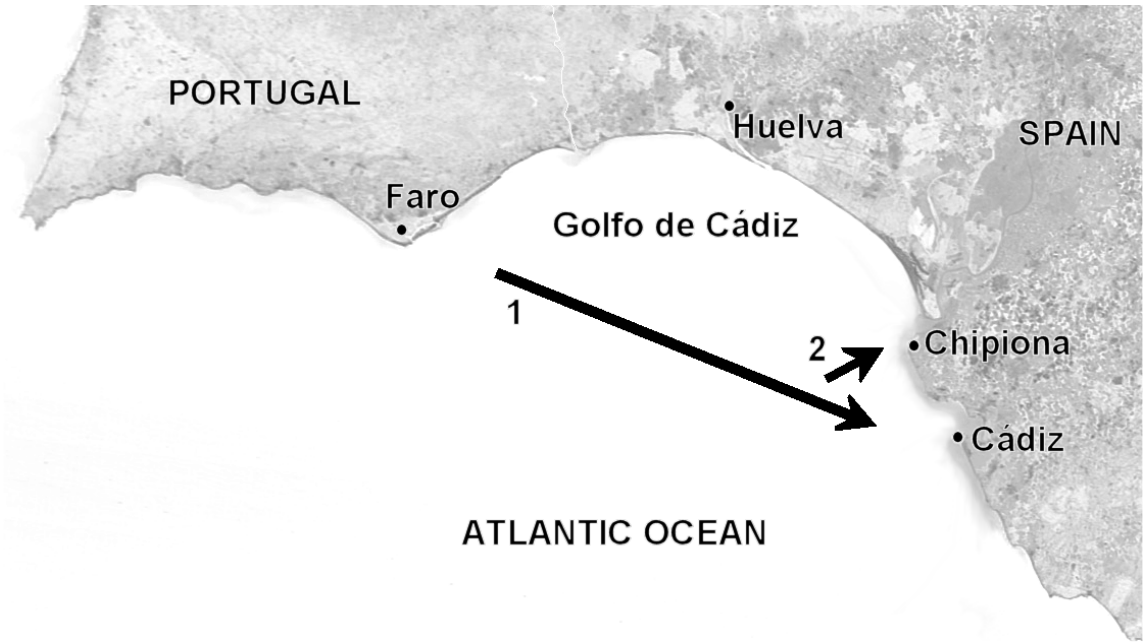

Figure 3. Projection on the ground of the atmospheric path of events SPMN050709 (1) and SPMN040710 (2).

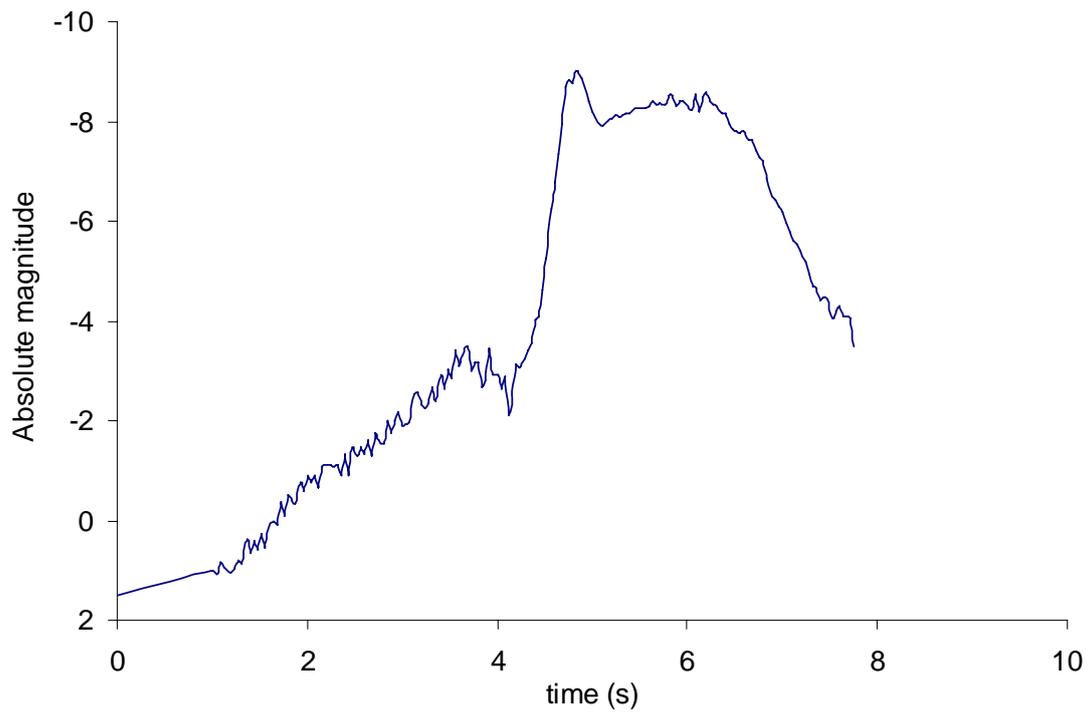

Figure 4. Lightcurve (absolute magnitude vs. time) of the SPMN050709 fireball.

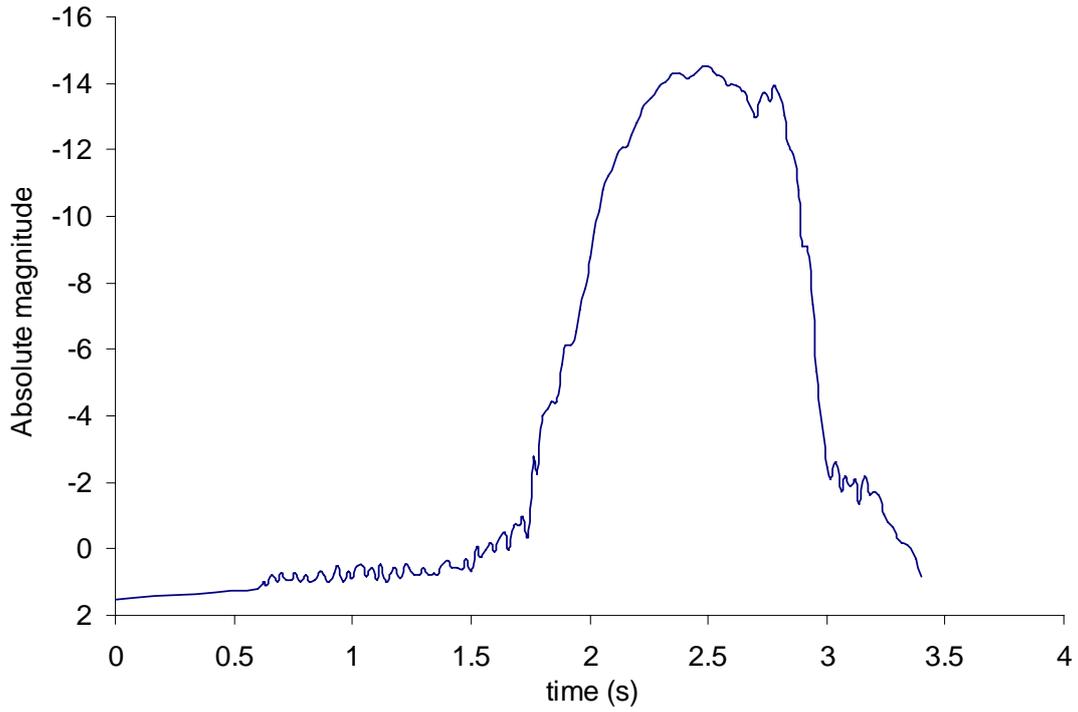

Figure 5. Lightcurve (absolute magnitude vs. time) of the SPMN040710 fireball.

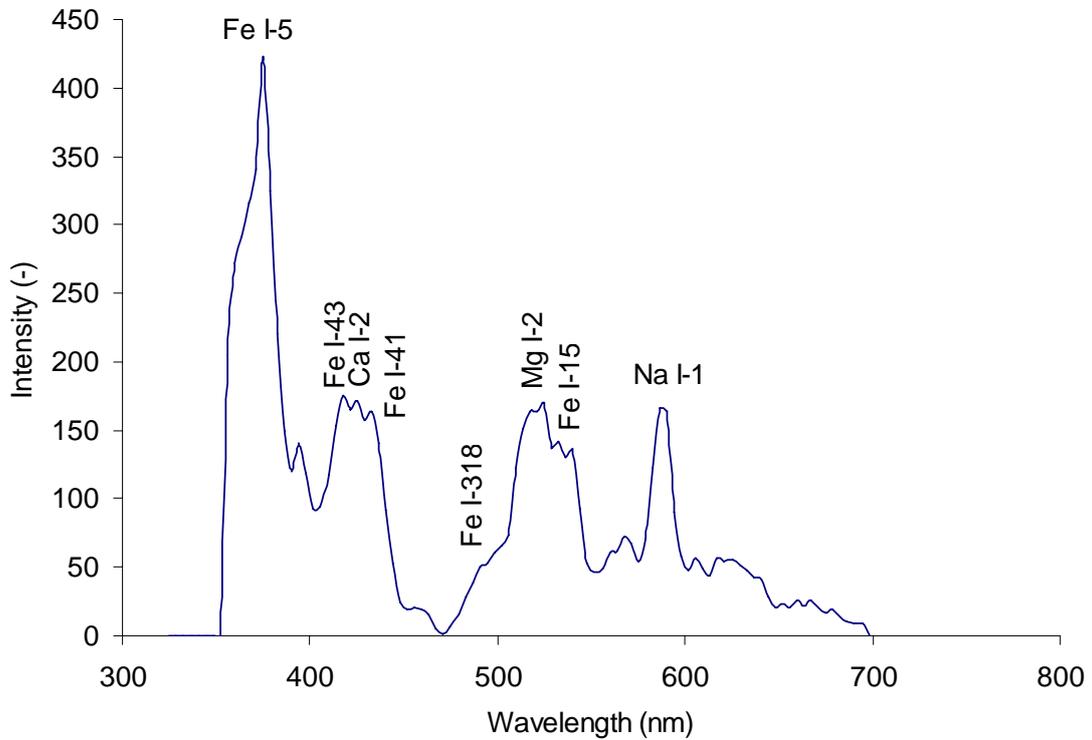

Figure 6. Calibrated emission spectrum recorded for the SPMN050709 fireball.
Intensity is expressed in arbitrary units.

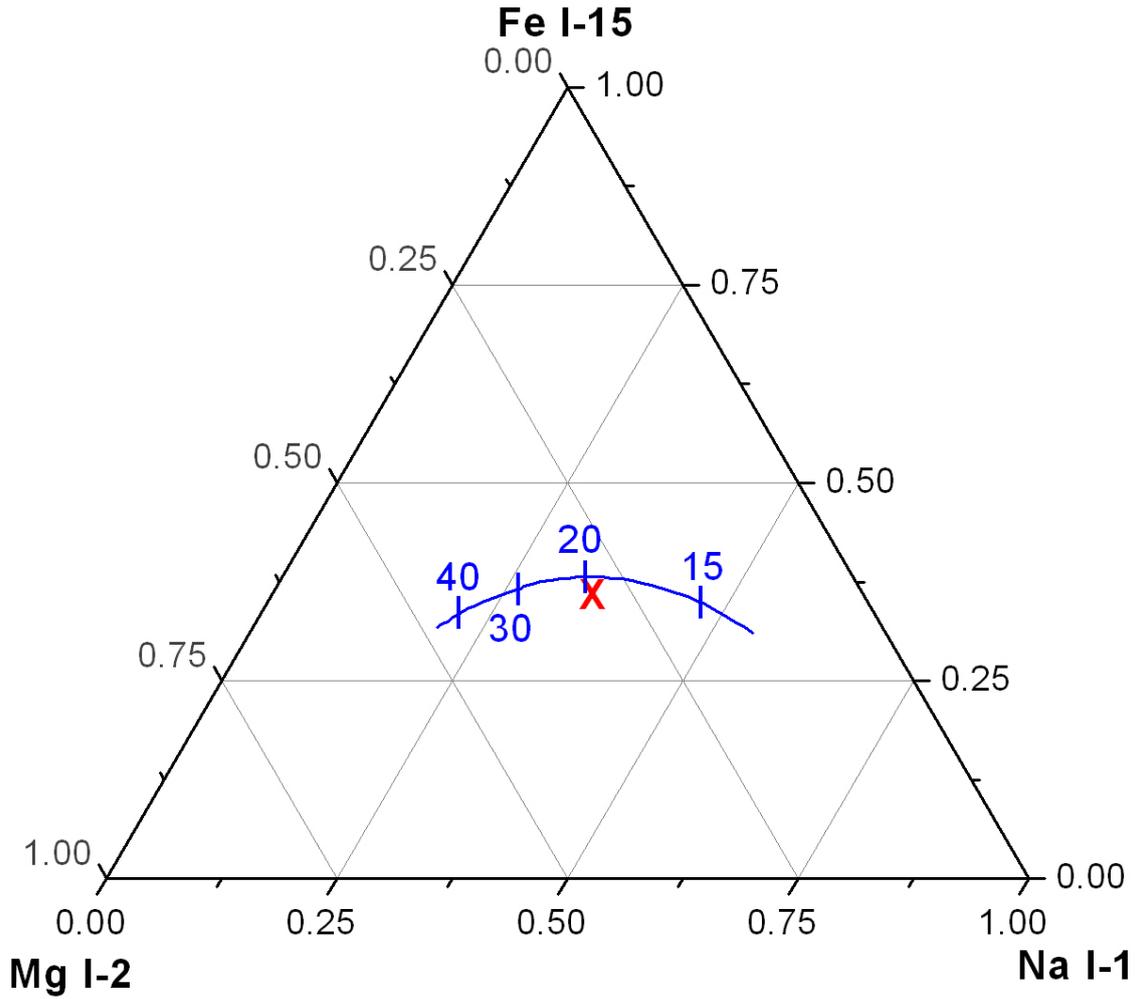

Figure 7. Expected relative intensity (solid line), as a function of meteor velocity (in km s$^{-1}$), of the Na I-1, Mg I-2 and Fe I-15 multiplets for chondritic meteoroids (Borovička et al. 2005). The cross shows the experimental relative intensity obtained for the SPMN050709 bolide.

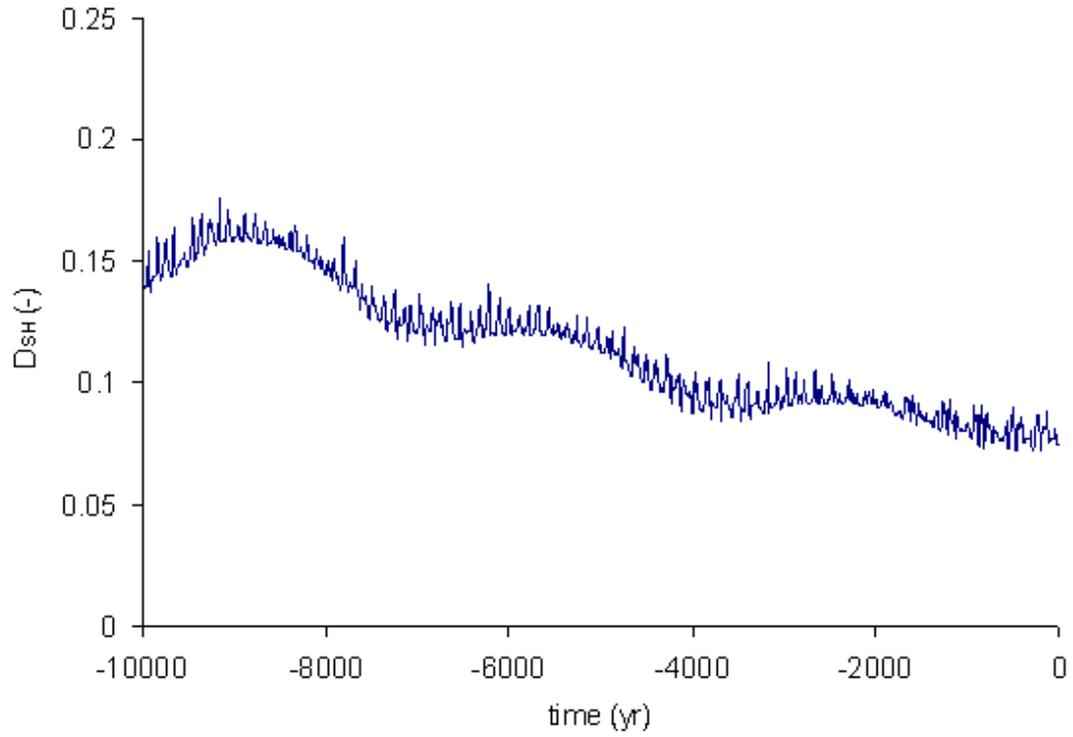

Figure 8. Evolution from present time of the $D_{SH}$ criterion for NEO 2007LQ19 and the averaged orbit of both fireballs.

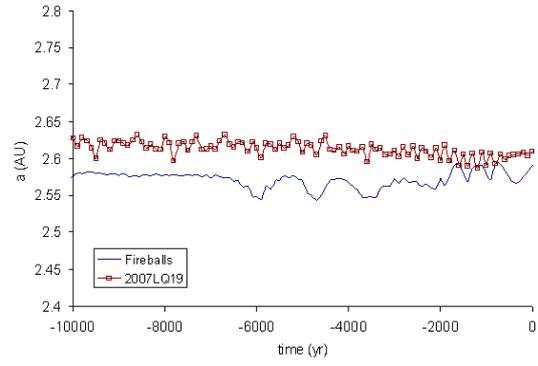

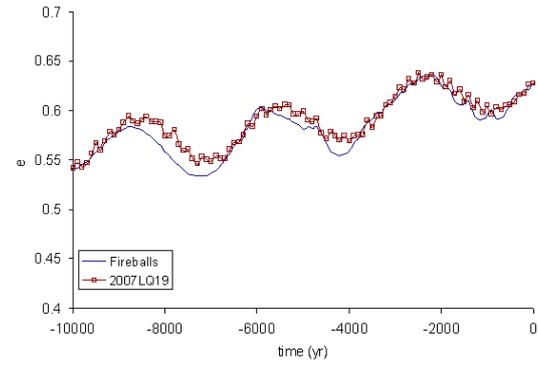

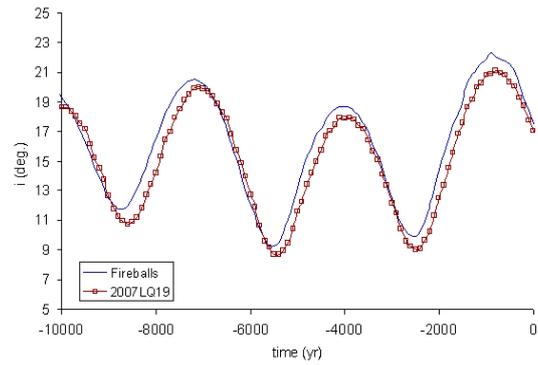

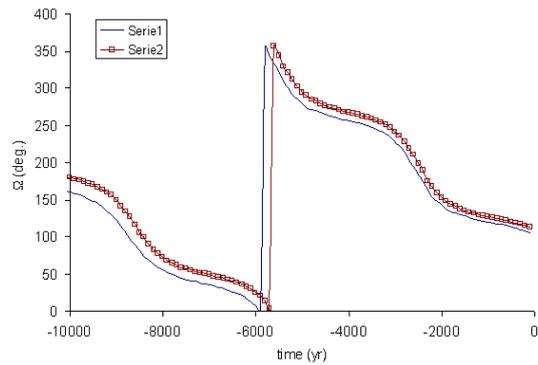

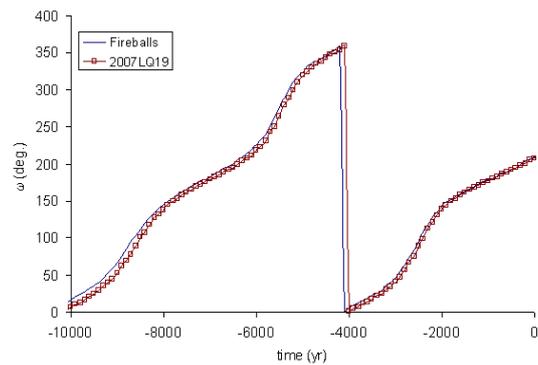

Figure 9. Evolution from present time of the orbital elements of NEO 2007LQ19 and the averaged orbit of both fireballs, where a is the semi-major axis, e the orbital eccentricity, i the inclination, $\Omega$ the longitude of the ascending node and $\omega$ the argument of perihelion.